\documentclass[jkps,preprint,fleqn,showpacs,showkeys]{revtex4}
\usepackage{graphicx}
\usepackage{amssymb}
\usepackage{amsmath}
\usepackage{bm}
\begin{document}
\newcommand{\be}{\begin{equation}}
\newcommand{\ee}{\end{equation}}
\newcommand{\bea}{\begin{eqnarray}}
\newcommand{\eea}{\end{eqnarray}}
\setcounter{page}{0}
\title[]{Curvature as a Measure of the Thermodynamic Interaction}
\author{Hernando \surname{Quevedo}}
\affiliation{Dipartimento di Fisica, Universit\`a di Roma ``La Sapienza", Piazzale Aldo Moro 5, I-00185 Roma, Italy;  
ICRANet, Piazza della Repubblica 10, I-65122 Pescara, Italy}
\email{quevedo@nucleares.unam.mx}
\thanks{on sabbatical leave from Instituto de Ciencias Nucleares, Universidad Nacional Aut\'onoma de M\'exico}
\author{Alberto S\'anchez}
\affiliation{
Departamento de Posgrado, CIIDET, AP 752, Quer\'etaro, QO 76000, Mexico} 
\email{asanchez@nucleares.unam.mx}
\author{Safia Taj}
\affiliation{
Dipartimento di Fisica, Universit\`a di Roma ``La Sapienza", Piazzale Aldo Moro 5, I-00185 Roma, Italy;  
ICRANet, Piazza della Repubblica 10, I-65122 Pescara, Italy}
\thanks{Permanent address: Center for Advanced Mathematics and Physics,
National University of Sciences and Technology,  Rawalpindi, Pakistan}
\email{safiataaj@gmail.com}
\author{Alejandro V\'azquez}
\affiliation{
Facultad de Ciencias, 
Universidad Aut\'onoma del Estado de Morelos, 
Cuernavaca, MO 62210, Mexico}
\email{alec_vf@nucleares.unam.mx}

\date[]{ January 2010}

\begin{abstract}
We present a systematic and consistent construction of geometrothermodynamics 
by using Riemannian contact geometry for the phase manifold and harmonic maps for the
equilibrium manifold. We present several metrics for the phase manifold that are invariant 
with respect to Legendre transformations and induce thermodynamic metrics on the 
equilibrium manifold. We review all the known examples in which the curvature of the 
thermodynamic metrics can be used as a measure of the thermodynamic interaction.

\end{abstract}


\maketitle

\section{INTRODUCTION}
\label{sec:int}
Differential geometry is a very important tool of modern science, 
especially mathematical physics, and it has many applications in physics,
chemistry and engineering. In particular, the four known interactions
of nature can be described in terms of geometrical concepts. Indeed, 
Einstein proposed the astonishing principle ``field strength = curvature" 
to understand the physics of the gravitational field (see, for instance, Refs. 1 and 2). 
In this case, 
curvature means the curvature of a Riemannian manifold. 
In general relativity, the connection involved 
is unique as a consequence of the assumption that the torsion tensor vanishes. 
The idea of this construction can be represented schematically as
\be
{\rm metric}\rightarrow {\hbox{\rm Levi-Civita connection}}\rightarrow {\hbox{\rm Riemann curvature = gravitational field strength}}.
\nonumber
\ee
The second  element of general relativity is Einstein's field equation
$
R_{\mu\nu}- \frac{1}{2} g_{\mu\nu} R = 8\pi T_{\mu\nu}
$,
which established for the first time the amazing principle ``geometry = energy." 
The conceptual fundamentals of this principle were very controversial; however, experimental 
evidence has shown its correctness, and even modern generalizations of Einstein's theory follow the same principle. 
On the other hand, because the field strength can be considered as a measure of the gravitational interaction, we conclude 
that the entire idea of general relativity can be summarized in the principle ``interaction = curvature." 

The discovery by Yang and Mills \cite{ym53} that the theory of electromagnetism can be described geometrically in terms of 
the elements of a principal fiber bundle constitutes an additional major achievement. The base manifold is in this case 
the Minkowski spacetime, the standard fiber is the gauge group $U(1)$, which 
represents the internal symmetry of electromagnetism,
and the connection across the fibers is a local cross-section, which takes values in the algebra of $U(1)$.   
This has opened the possibility of fixing the background metric 
in accordance with the desired properties of the base manifold and selecting different connections as local cross-sections 
of the principal fiber bundle. This interesting geometrical approach constitutes the base for constructing the modern 
gauge theories that are used to describe the physics of the electromagnetic, the weak, and the strong interactions. 
This construction can be represented schematically as 
\be
\begin{array}[c]{ccccc}
              &                &  U(1)-{\rm connection} & \rightarrow & U(1)-{\rm curvature} = {\hbox{\rm electromagnetic interaction}}  \\
   & \nearrow       &  \\
{\hbox{\rm Minkowski metric}}  & \rightarrow    &   SU(2)-{\rm connection} & \rightarrow & SU(2)-{\rm curvature} = {\hbox{\rm weak interaction}}\\
             & \searrow       & \\
              &                &  SU(3)-{\rm connection} & \rightarrow & SU(3)-{\rm curvature} = {\hbox{\rm strong interaction.}} \nonumber
\end{array}
\ee
We conclude that the principle ``curvature = interaction" holds for all known forces of nature. 

Consider now the case of a thermodynamic system.
In very broad terms, one can say that in a thermodynamic system, all the known forces act among the particles that constitute the system.  
Due to the large number of particles involved in the system, only a statistical approach is possible, from which average values for
the physical quantities of interest are derived. The question arises whether it is also possible to find a geometric construction for which the 
principle ``curvature = thermodynamic interaction" holds. We will see in the present work that the formalism of geometrothermodynamics (GTD) \cite{quev07}
satisfies this condition. First, we must mention that our interpretation of the thermodynamic interaction is based upon a statistical approach 
to thermodynamics in which all  the properties of the system can be derived from the explicit form of the corresponding Hamiltonian \cite{greiner},
and in which the interaction between the particles of the system is described by the potential part of the Hamiltonian. Consequently, if the potential vanishes, we say that the system has a zero thermodynamic interaction. 

In this work, we present the formalism of GTD by using  Riemannian contact geometry for the definition of the thermodynamical phase 
manifold and harmonic maps for the definition of the equilibrium manifold. We will see that this approach allows us 
to interpret any thermodynamic system  as a hypersurface in the equilibrium space completely determined by the
field theoretical approach of harmonic maps. 
This paper is organized as follows: In Section \ref{sec:rcg}, we introduce the main concepts of Riemannian contact geometry 
that are necessary to define the phase manifold. Section \ref{sec:har} is dedicated to the application of harmonic maps
to obtain different aspects of GTD, like the equilibrium manifold and its geodesics. Furthermore, in Section \ref{sec:res},
we present a series of thermodynamic systems described by Riemannian manifolds whose curvature can be interpreted as 
a measure of thermodynamic interaction. Finally, Section \ref{sec:con} is devoted
to discussions of our results and suggestions for further research.
Throughout this paper, we use units in which $G=c=k_{_B}=\hbar =1$.

\section{Riemannian contact geometry} 
\label{sec:rcg}
The main element of GTD is the thermodynamic phase manifold, which is a Riemannian contact manifold 
whose contact structure and metric are invariant with respect to Legendre transformations. First, we 
consider a $(2n+1)-$dimensional differential manifold ${\cal T}$ and its tangent manifold $T{\cal T}$. 
Let ${\cal V}\subset T{\cal T}$ be a field of hyperplanes on ${\cal T}$. It can be shown that 
${\cal V} = \ker \Theta$, i.e.,  the kernel of a non-vanishing differential 1-form $\Theta$. 
If the Frobenius integrability condition $\Theta\wedge d \Theta =0$ is satisfied, the hyperplane field ${\cal V}$ is
completely integrable. On the other hand, if   $\Theta \wedge d \Theta \neq 0$, then ${\cal V}$ is non-integrable. 
In the limiting case $\Theta \wedge (d \Theta)^n \neq 0$, the hyperplane field ${\cal V}$ becomes maximally non-integrable
and  is said to define a contact structure on ${\cal T}$. 
The pair $({\cal T},{\cal V})$ is usually known as a contact manifold \cite{handbook}
and is sometimes denoted as $({\cal T},\Theta)$
to emphasize the role of the contact form $\Theta$. 

Let $G$ be a non-degenerate metric on ${\cal T}$. The set $({\cal T}, \Theta, G)$ defines a Riemannian contact manifold.
Notice that the contact manifold $({\cal T}, \Theta)$ is (almost) uniquely defined in the following sense: The condition
$\Theta \wedge (d\Theta)^n \neq 0$ is independent of $\Theta$; in fact, it is a property of ${\cal V}=\ker\Theta$.
If another 1-form $\Theta'$ generates the same ${\cal V}$, it must be of the form 
$\Theta' = f \Theta$, where $f: {\cal T} \rightarrow \mathbb{R}$ is a smooth non-vanishing function. The Riemannian metric $G$, instead,
is completely arbitrary. We will use this freedom to select only those metrics that are invariant under Legendre transformations.

To introduce Legendre transformations in a general fashion, 
let us choose the coordinates of ${\cal T}$ as $Z^A =\{\Phi, E^a, I ^a\}$ with $a=1,...,n$, and $A=0,1,...,2n$. Here, $\Phi$ represents
the thermodynamic potential used to describe the system whereas the coordinates $E^a$ correspond to the extensive variables and $I^a$ to the
intensive variables.  Notice that in the manifold ${\cal T}$, all the coordinates $\Phi$, $E^a$ and $I^a$ must be completely independent;
thus, thermodynamic systems cannot be described in ${\cal T}$. A Legendre transformation is defined as \cite{arnold}
\be
\{Z^A\}\longrightarrow \{\widetilde{Z}^A\}=\{\tilde \Phi, \tilde E ^a, \tilde I ^ a\}\ ,
\ee
\be
 \Phi = \tilde \Phi - \delta_{kl} \tilde E ^k \tilde I ^l \ ,\quad
 E^i = - \tilde I ^ {i}, \ \  
E^j = \tilde E ^j,\quad   
 I^{i} = \tilde E ^ i , \ \
 I^j = \tilde I ^j \ ,
 \label{leg}
\ee
where $i\cup j$ is any disjoint decomposition of the set of indices $\{1,...,n\}$,
and $k,l= 1,...,i$. In particular, for $i=\{1,...,n\}$ and $i=\emptyset$, we obtain
the total Legendre transformation and the identity, respectively. In these particular
coordinates, the contact 1--form can be written as 
\be
 \Theta= d\Phi - \delta_{ab} I^a d E^b \ ,\quad \delta_{ab}={\rm diag} (1,1,...,1)\ ,
\label{gibbs}
\ee
an expression that is manifestly invariant with respect to the Legendre 
transformations in Eq. (\ref{leg}). Consequently, the contact manifold $({\cal T}, \Theta)$ is a Legendre
invariant, as will be the Riemannian contact manifold $({\cal T}, \Theta,G)$, if we demand 
Legendre invariance of the metric $G$. 

Any Riemannian contact manifold $({\cal T}, \Theta,G)$ whose components are Legendre invariant is 
called a thermodynamic {\it phase manifold} and constitutes the starting point for a description 
of thermodynamic systems in terms of geometric concepts. We would like to emphasize the fact that Legendre invariance
is an important condition that guarantees that the description does not depend on the choice of the 
thermodynamic potential, a property that is essential in ordinary thermodynamics.

As mentioned before, the only freedom in the construction of the phase manifold is in the choice of the metric $G$.
Legendre invariance implies a series of algebraic conditions for the metric components $G_{AB}$ 
\cite{quev07}, and it can be shown that these conditions are not trivially satisfied. For instance, 
a straightforward computation shows that the flat 
metric $G=\delta_{AB}dZ^A dZ^B$ is not invariant with respect to the Legendre transformations in Eq. (\ref{leg}). 
It then follows that the phase space is necessarily curved. We performed a detailed analysis
of the Legendre invariance conditions and found that the metric 
\be
\label{ginv2}
G=\left(d\Phi - I_a dE^a\right)^2 + \Lambda\, (E_a I_a)^{2k+1} d E^a d I^a \ ,\quad E_a =\delta_{ab}E^b\ ,\quad
I_a =\delta_{ab} I ^b\ ,
\ee
where $\Lambda$ is an arbitrary real constant and $k$ is an integer, is invariant with respect to partial and total Legendre transformations. 
To our knowledge, this is the most general metric satisfying the conditions of Legendre invariance. 
The corresponding scalar curvature
\be
R= \frac{2}{\Lambda^2} \left\{\left[ \sum_{a=1}^n (E_aI_a)^{-2k-1}\right]^2 
- 3 \sum_{a\neq b}^ n (E_a I_a E_b I_b)^{-2k-1}\right\} 
\ee
shows that the manifold is curved in general.

Furthermore, the phase manifold metric
\begin{equation}
\label{ginv1}
G = (d\Phi - I_{a}dE^{a})^{2} + \Lambda
(E_{a}I^{a})^{2k+1} (\chi_{bc}dE^{b}dI^{c})\ ,
\end{equation}
where $\Lambda$ is constant, $k$ is an integer,  and $\chi_{ab}$ is a constant diagonal tensor, satisfies the conditions that follow from 
a total Legendre transformation, Eq. (\ref{leg}). The corresponding curvature is rather cumbersome and cannot be written in a compact form; however, an inspection of its explicit form shows that it is always different from zero. 

The metrics in Eqs. (\ref{ginv2}) and (\ref{ginv1}) are the most general Legendre invariant metrics we have found so far 
and contain other known metrics as particular cases \cite{quev07}. Legendre transformations impose, in general, very strong 
conditions on the components $G_{AB}$; indeed, Eq. (\ref{leg}) shows that such a transformation can change an extensive variable
to the negative of the corresponding intensive variable. This implies that only very specific combinations of extensive and intensive
variables can be invariant under Legendre transformations.

\section{Harmonic maps}
\label{sec:har}

Consider two (pseudo)-Riemannian manifolds $(M,\gamma)$ and $(M', \gamma')$ of dimension
$m$ and $m'$, respectively. Let the base manifold $M$ be coordinatized by $x^\alpha$ 
($\alpha,\beta,\gamma,... =1,2,..., m$), and $M'$ by $x'^\mu$ 
($\mu,\nu,\lambda,... = 1,2,..., m'$),
so that the metrics on $M$ and $M'$ can be, in general, smooth functions of the 
corresponding coordinates, i.e., $\gamma=\gamma(x)$ and $\gamma'=\gamma'(x')$.  
A harmonic map is a smooth map $\varphi: M \rightarrow M'$, or in coordinates
$\varphi: x \longmapsto x'$ so that $x'$ becomes a function of $x$. The $x'$s  
satisfy the field equations following from the action \cite{misner}
\be
S = \frac{1}{2}\int d^m x \sqrt{|\det(\gamma)|}\ \gamma^{\alpha\beta}(x)\ \frac{\partial x'^\mu}{\partial x^\alpha}
\frac{  \partial x'^\nu}{ \partial x^\beta}
\gamma' _{\mu\nu}(x')  \ ,
\label{acts}
\ee
which sometimes is called the ``Dirichlet energy functional" of the harmonic map $\varphi$.
The straightforward variation of $S$ with respect to $x'^\mu$ leads 
to the field equations
\be
\frac{1}{\sqrt{|\det(\gamma)|}} \frac{\partial}{\partial x^\beta}
\left(\sqrt{|\det(\gamma)|}\gamma^{\alpha\beta} \frac{\partial x'^\mu}{\partial x^\alpha} \right) 
+ \Gamma^\mu_{\ \nu\lambda} \ \gamma^{\alpha\beta}\frac{ \partial x'^\nu}{\partial x^\alpha}
\frac{ \partial x'^\lambda}{\partial x^\beta} 
 = 0 \ ,
\label{moteq}
\ee
where $\Gamma^\mu_{\ \nu\lambda}$ are the Christoffel symbols associated with the 
metric $\gamma'_{\mu\nu}$ of the target manifold $M'$. If $\gamma'_{\mu\nu}$ is a flat metric,
one can choose Cartesian-like coordinates such that $\gamma'_{\mu\nu}=\chi_{\mu\nu} =
{\rm diag}(\pm 1, ..., \pm 1)$, the field equations become linear, and the
 harmonic map is linear. 
In the following subsections, we will show that harmonic maps are the correct 
mathematical tool to investigate the properties of the phase manifold and its submanifolds, which 
contain information on the physical states of thermodynamic systems. 

\subsection{Geodesics of the Phase Manifold}
\label{ssec:geop}
Consider a base manifold with $\dim(M) = 1$ and identify the target manifold with 
the thermodynamic phase manifold $({\cal T},\Theta,G)$. Then, the field 
equations, Eqs. (\ref{moteq}), reduce to the geodesic equations
\be
\frac{d^2Z^A}{d\lambda^2} + \Gamma^A_{\ BC} \frac{dZ^B}{d\lambda} \frac{dZ^C}{d\lambda} =0 \ ,
\ee
where $\lambda$ is an affine parameter and $\Gamma^A_{\ BC}$ are the Christoffel symbols of the phase manifold metric $G$.
Since any Legendre invariant $G$ has a non-zero curvature, these geodesic equations are highly non-linear and difficult to
solve in general. Even special cases of the known metrics in Eqs. (\ref{ginv2}) and (\ref{ginv1}) require a detailed analysis that
is beyond the scope of the present work. Preliminary results indicate that the geodesics of the phase manifold represent 
families of thermodynamic systems that can be investigated in the context of GTD.

\subsection{Equilibrium Manifold}
\label{ssec:eqm}
Consider the harmonic map $\varphi: {\cal E}\rightarrow {\cal T}$, where ${\cal E}$ is a subspace of the phase manifold $({\cal T},\Theta,G)$ 
and $\dim({\cal E}) = n$. For the sake of concreteness, let us assume that the extensive variables $\{E^a\}$ are the coordinates of  ${\cal E}$.
Then, in terms of coordinates, the harmonic embedding map reads $ \varphi :  \{E^a\} \longmapsto \{Z^A(E^a)\}=\{\Phi(E^a), E^a, I^a(E^a)\}$. Moreover, 
the pullback $\varphi^*$ of the harmonic map induces canonically a metric $g$ on  ${\cal E}$ by means of 
\be 
g=\varphi^* (G)\ ,\quad {\rm i.e.}\quad 
g_{ab} = \frac{\partial Z^A}{\partial E^a} \frac{\partial Z^B}{\partial E^b} G_{AB} 
= Z^A_{,a} Z^B_{,b} G_{AB} \ .
\ee
If we assume that the metric $\gamma$ of the base manifold coincides with the induced metric $g$, the action in Eq. (\ref{acts}) reduces to 
\be
S= \frac{n}{2}\int d^n E \sqrt{|\det(g)|}\ ,
\label{ng}
\ee
and the field equations become
\be
\frac{1}{\sqrt{|\det(g)|}}\left(\sqrt{|\det(g)|}\,\, g^{ab}Z^A_{,a}\right)_{,b} + 
\Gamma^A_{\ BC} Z^B_{,b}Z^C_{,c} g^{bc} =0 \ .
\label{meng}
\ee
The action in Eq. (\ref{ng}) corresponds to the volume element of the submanifold ${\cal E}\subset {\cal T}$; 
consequently, the field
equations in Eqs. (\ref{meng}) represent the condition for ${\cal E}$ to be an extremal hypersurface in the phase manifold
\cite{vqs09}. If, furthermore, the harmonic map satisfies the condition $\varphi^*(\Theta)=0$, the pair $({\cal E},g)$ is called
an {\it equilibrium manifold}. The last condition is equivalent to
\begin{equation}
 d\Phi
=I_{a}dE^{a}\ , \quad \frac{\partial \Phi}{\partial E^{a}}=I_{a}\ .
\label{firstlaw}
\end{equation}
The first of these equations corresponds to the first law of
thermodynamics whereas the second one is usually known as the
condition for thermodynamic equilibrium \cite{callen}. 
We conclude that the harmonic map $\varphi:{\cal E} \rightarrow {\cal T}$ 
defines the equilibrium manifold $({\cal E},g)$ as an extremal submanifold of the phase manifold $({\cal T},\Theta,G)$ in which 
the first law of thermodynamics and the equilibrium conditions for a given system with fundamental equation $\Phi=\Phi(E^a)$ hold.

\subsection{Geodesics of the Equilibrium Manifold}
\label{ssec:geoe}
Consider a base manifold with $\dim(M)=1$ and identify the target manifold $M'$ with the equilibrium manifold $({\cal E},g)$
defined above. Then, the field equations reduce to the geodesic equations in the equilibrium manifold:
\be
\frac{d ^2E^a}{d\tau^2} + 
\Gamma^a_{\ bc} \frac{dE^b}{d\tau} \frac{dE^c}{d\tau} = 0 \ ,
\label{geo1}
\ee
where $\Gamma^a_{\ bc}$ are the Christoffel symbols of the thermodynamic metric $g$, and $\tau$ is an arbitrary affine parameter
along the geodesic. Because the points of the equilibrium manifold represent equilibrium states in which the thermodynamic system 
can exist, a geodesic in $({\cal E},g)$ represents a quasi-static process along which the system can evolve. This has been shown 
explicitly in the case of an ideal gas \cite{qsv09}.

\section{Examples}
\label{sec:res}
The harmonic maps presented in the last sections allow us to define geometric structures 
in an invariant way. In particular, the curvature of the thermodynamic metric $g$ should 
represent the thermodynamic interaction independently of the thermodynamic potential. In fact,
this is not a trivial condition from a geometric point of view. In particular, a geometric analysis 
of black--hole thermodynamics by using metrics introduced {\it ad hoc} in the equilibrium manifold
leads to contradictory results \cite{aman06a,caicho99,curflat}. Using the induced thermodynamic metric
$g$ as defined in Section \ref{ssec:eqm}, by means of the pullback of the harmonic map, the results are
consistent and invariant. For instance, in the particular case of the phase metric in Eq. (\ref{ginv1}) with $k=0$ and $\chi_{ab}=
\eta_{ab}={\rm diag}(-1,1,...,1)$, the induced thermodynamic metric 
\be
g=\varphi^*(G)= \left(E^{c}\frac{\partial{\Phi}}{\partial{E^{c}}}\right)
\left(\eta_{ab}\delta^{bc}\frac{\partial^{2}\Phi}{\partial {E^{c}}\partial{E^{d}}} dE^a dE^d \right)\ 
\label{gdown1}
\ee
has been applied to a large class of black--hole configurations
in the following theories \cite{aqs08,qs08,qs09a,qs09b,qt09,akbar10}: 2--dimensional dilaton gravity, 
3--dimensional Einstein gravity, 4-- and higher--dimensional 
Einstein-Maxwell theory with a cosmological constant, and 5--dimensional Einstein-Gauss-Bonnet theory with electromagnetic
and Yang-Mills fields. As a general result, we find that the curvature of the equilibrium manifold 
is non-vanishing, that it can be used as a measure of the thermodynamic interaction, and that it diverges
at those points where second--order phase transitions occur.

From the phase manifold metric in Eq. (\ref{ginv2}), we obtain the thermodynamic metric  
\be
g= \Lambda\left(E_a\frac{\partial\Phi}{\partial E^a}\right)^{2k+1} 
\frac{\partial^2\Phi}{\partial E^b \partial E^c}\delta^{ab} dE^a dE^c
\ ,
\label{gdown2}
\ee
which has been used to describe GTD of ordinary thermodynamic systems, like an ideal gas and its non-interacting generalizations, 
a van der Waals gas, and a 1--dimensional Ising model. We have shown that the arbitrary constant $k$ can be chosen such that
the field equations are satisfied. Moreover, we found that the curvature of the equilibrium manifold
vanishes only in the case of non-interacting systems, that it is non-zero for interacting systems, and that it diverges at those
points where first--order phase transitions occur.

\section{Conclusions}
\label{sec:con}

In this paper, we showed that harmonic maps play an important role in the formalism of geometrothermodynamics (GTD). 
They can be used to derive  geodesic equations in different spaces and to introduce in a consistent 
and invariant way the concept of an equilibrium manifold. It turns out that for a given fundamental 
equation of the form $\Phi=\Phi(E^a)$, GTD provides an invariant approach to construct 
the corresponding equilibrium manifold whose points represent equilibrium states. 
The harmonic map, which determines 
the equilibrium manifold, also generates  a system of differential equations 
that determine extremal hypersurfaces in the phase manifold. 
This construction allows us to investigate the properties of the 
curvature of the equilibrium manifold and to propose it as an invariant measure 
of the thermodynamic interaction. We presented all the examples for which  
we have shown that the thermodynamic curvature not only measures the interaction 
of the thermodynamic system but also becomes singular 
at those points where phase transitions occur. Thus, GTD represents
an invariant geometric formalism of standard thermodynamics that resembles 
the famous principle ``curvature =  interaction," which is  valid for all known forces
of nature.

\begin{acknowledgements}

H. Quevedo and S. Taj would like to thank ICRANet for support. 
\end{acknowledgements}



\begin{references}

\bibitem{mtw} C. W. Misner, K. S. Thorne, and J. A. Wheeler, {\em Gravitation} (W. H. Freeman, San Francisco, 1973).

\bibitem{frankel} T. Frankel, {\em The Geometry of Physics: An Introduction} (Cambridge University Press, Cambridge, UK, 1997).

\bibitem{ym53} C. N. Yang and R. L.  Mills, Phys. Rev. {\bf 96}, 191 (1954).

\bibitem{quev07} H. Quevedo,  J. Math. Phys. {\bf 48}, 013506 (2007). 

\bibitem{greiner} W. Greiner, L. Neise and H. St\"ocker, {\it Thermodynamics and Statistical Mechanics} (Springer Verlag, New York, 
1995).

\bibitem{handbook} F. Dillen and L. Verstraelen, {\it Handbook of Differential Geometry} (Elsevier B. V., Amsterdam, 2006).


\bibitem{arnold} V. I. Arnold, {\it Mathematical Methods of Classical Mechanics}
(Springer Verlag, New York, 1980).

\bibitem{misner} C. W. Misner, Phys. Rev. D {\bf 18}, 4510 (1978). 
 
\bibitem{vqs09} A. V\'azquez, H. Quevedo, and A. S\'anchez, arXiv:hep-th/0805.4819 (2009).


\bibitem{callen} H. B. Callen, {\it Thermodynamics and an Introduction to 
Thermostatics} (John Wiley \& Sons, Inc., New York, 1985).

\bibitem{qsv09} H. Quevedo, A. S\'anchez and A. V\'azquez, arXiv:math-phys/0811.0222 (2009).



\bibitem{aman06a} J. \AA man, I. Bengtsson, and  N. Pidokrajt, 
Gen. Rel. Grav. {\bf 38}, 1305 (2006). 



\bibitem{caicho99} R. Cai and J. Cho, 
Phys. Rev. D {\bf 60}, 067502 (1999). 


\bibitem{curflat} B. Mirza and M. Zamaninasab,  J. High Energy Phys., 0706:059 (2007).    

\bibitem{aqs08} J. L. \'Alvarez, H. Quevedo, and A. S\'anchez,  Phys. Rev. D {\bf 77}, 084004 (2008). 

\bibitem{qs08} H. Quevedo and A. S\'anchez,  JHEP {\bf 09}, 034 (2008). 

\bibitem{qs09a} H. Quevedo and A. S\'anchez, 
Phys. Rev. D {\bf 79}, 024012 (2009). 

\bibitem{qs09b} H. Quevedo and A. S\'anchez, 
Phys. Rev. D. {\bf 79}, 087504 (2009).
 


\bibitem{qt09} H. Quevedo and S. Taj, {\it Geometrothermodynamics of Higher Dimensional Black Holes in
Einstein-Gauss-Bonnet theory}, (2010),  in preparation. 


\bibitem{akbar10} M. Akbar, H. Quevedo,  K. Saifullah, A. S\'anchez, and S. Taj {\it Thermodynamic Geometry of
Charged Rotating BTZ Black Holes}, (2010),  in preparation.




\end{references}
\end{document}